\documentstyle[a4,11pt,graphicx]{article}
\title{The Ages, Masses, Evolution and Kinematics of Mira Variables}
\author{Michael Feast\\
Astronomy Department, University of Cape Town}
\date{   }
\begin{document}
\maketitle
\begin{abstract}
Evidence on the ages and masses of Mira variables is reviewed.
Period increases with increasing initial mass. 
Miras of $\log P \sim 3.0$ have initial masses near $4 M_{\odot}$.
It is suggested that the apparent gap in the LMC Mira PL relation
at about this period
may be due to the onset of hot bottom  burning  and that 
this adds $\sim15$ to 20 percent to the stellar energy production.
Shorter period HBB stars are probably overtone pulsators. T Lep
may be an example of cool bottom processing.
\end{abstract}

\section{Introduction}
Mira variables are at the final stage of AGB star evolution and their
observation is of particular importance since theory has so far been 
unable to make firm quantitative predictions regarding several aspects
of this phase.
They are also
the brightest objects (particularly in the infrared) in old and
intermediate age populations. They are thus important for the
study of the distribution and kinematics of old and intermediate populations
in our own and other galaxies. Several examples of progress in both these
areas using the Japanese-South African IRSF-Sirius combination 
have been given at this meeting (e.g.
\cite{Matsunaga-2008} \cite{Ita-2008} \cite{Menzies-2008}). 
The aim of the present paper is
to summarize some of the relevant observational data 
on the ages, masses, state of evolution and kinematics of
these objects and to draw tentative
conclusions which further work may confirm or falsify. 

\section{The PL Relation and the Miras above it}
 Miras in the LMC are found to define a narrow 
period-luminosty (PL) relation at $K$ extending from 
$\log P \sim 2.1$ to 2.6
\cite{Feast-1989}. Oxygen-rich (O-Miras) and carbon-rich (C-Miras)
Miras lie together on this relation. A similar relation holds in
$M_{bol}$ (see e.g. \cite{Whitelock-2008a}). In this case there may be 
a slightly different relation for
O-Miras than for C-Miras. However, in the range
$\log P \sim 2.6$ to 2.8, the LMC Miras lie systematically above a linear
extrapolation of the PL in either $K$ or $M_{bol}$. 
The VERA parallax of UX Cyg 
shows that such stars exist in our own Galaxy
\cite{Kurayama-2005} \cite{Whitelock-2008}.
In 1989 \cite{Feast-1989} 
the reason for these 
``above-PL" stars was not known and it was suggested that they might
evolve into OH/IR stars at much longer periods
(i.e. that they were not yet, like ``true" Miras, at the end of
their AGB evolution). Alternatively,
Hughes and Wood \cite{Hughes-1990}, who found more objects in the 
$\log P \sim 2.6$ to 2.8
range, suggested that 
there was a real steepening of the PL slope at about
$\log P = 2.6$. However, all these early studies relied on optically
selected samples. The matter was very considerably clarified by
Whitelock et al. \cite{Whitelock-2003}. They studied objects in the LMC 
obscurred by
circumstellar shells and found Miras with periods in the range 
$\log P \sim 2.6$ to 3.2 
which lay on an extrapolation of the 
$M_{bol}$  PL relation at shorter periods.
They also
found that Miras in which lithium lines had been detected \cite{Smith-1995}
and which are thought to be Hot Bottom Burning (HBB) stars 
lay in the above-PL group, suggesting that all the above-PL Miras might be
HBB stars.
HBB is believed to be an important process in
the late evolution of stars of mass above about 3 or $4 M_{\odot}$. Such stars
are expected to dredge up and eject helium-rich material. This
is of current relevance for our understanding of globular cluster evolution.
If an earlier generation of intermediate mass AGB stars in a cluster enriched
the interstellar medium of the cluster in helium, this might explain
the helium-rich sequences found in some clusters. The matter remains
controversial and Renzini \cite{Renzini-2008}, who has summarized the present 
position,
points out that much depends on the details of the way HBB operates
and these can only be found observationally. It is evident therefore
that identifying and studying HBB Miras is a matter of particular interest. 

\section{The ages and initial masses of Miras}
 Galactic globular clusters contain O-Mira variables in the range
$\log P \sim 2.0-2.5$. These lie on the PL relation \cite{Feast-2002}.
There is also a rather clear 
period-metallicity relation \cite{Feast-2000}
\footnote{There may be complications to this in the case of
clusters showing the second parameter effect (e.g. NGC 6441)
\cite{Matsunaga-2007}.}.
Evidently these Miras all have low initial masses. It is possible that there
might be a general (slight) increase of initial mass with [Fe/H]. Whether
this is so or not depends both on the adopted stellar models and the
relative ages of clusters of different metallicity and thus remains uncertain.
Since the change of mass with period for the cluster stars must in any case be
small, the range in period must reflect an increase in stellar radius
caused by increasing [Fe/H]. 
 It is well known that there is a dependence of O-Mira kinematics on their
periods \cite{Feast-1963}. This allows one to estimate ages as a function of
period by comparison of O-Mira velocity dispersions \cite{Feast-2006} 
with relations between age and velocity dispersion
for stars in the solar neighbourhood \cite{Holmberg-2007}.
The shortest period group 
($\log P \sim 2.3$) are evidently globular-cluster-like and will
have initial masses of $<1 M_{\odot}$. The bulk of the Galactic Miras have
$\log P \sim 2.5$ and are $\sim 7$Gyr old. The group with $\log P \sim 2.65$
(which might contain some above-PL stars) is $\sim 3$Gyr old. 
Evidently these stars are of low initial masses though this increases
with increasing period.
Assuming that the OH/IR Mira with $\log P =3.107$ is a
member of the LMC cluster HS327
\cite{Van Loon-2001}, its initial mass is about
$4 M_{\odot}$ . Note that as discussed earlier \cite{Feast-2006} 
\cite{Feast-2007} the
O-Miras allow one to distinguish between two alternative interpretations
of Galactic kinematics. 

Galactic C-Miras are mostly confined to the longer periods. Their kinematics
at a mean $\log P =2.717$
\cite{Feast-2006} indicates a mean age of $\sim 1.8$Gyr and an initial
mass of $\sim 1.8 M_{\odot}$. There is an 
indication that, as for O-Miras, 
period increases with decreasing age.
At a given period the age of a C-Mira may be slightly
less than that of an O-Mira, but this is uncertain. The discovery 
\cite{Nishida-2000} of three clusters in the Magellanic Clouds containing
C-Miras was a major breakthrough in age/mass derivation. These 
stars which lie on the PL also have 
a mean age $\sim 1.8$Gyr 
(mean $\log P = 2.689$). Van Loon et al.\cite{Van Loon-2003} suggest that a
C-Mira with $\log P = 2.833$ belongs to the LMC cluster KMHK
with an age of $\sim 1.0$ Gyr and an initial mass of $\sim 2.2 M_{\odot}$,
consistent with the gradual increase of initial
mass as one moves up the PL.

\section{Hot Bottom Burning Stars}  
 Taking together the results for samples of LMC Miras selected optically
\cite{Feast-1989} and from infrared properties \cite{Whitelock-2003}
one finds the following for stars on or near the PL: (1) Between $\log P 
\sim 2.1$ and 2.6 there is a mixture of O and C-Miras. (2) Between  
$\log P \sim 2.6$ and 3.0 the stars are nearly all C- Miras. There is only
one O-Mira in this range and it is at the long period end ($\log P = 2.95$).
(3) There seems to be a rather conspicuous gap with no Miras between
$\log P = 2.97$ and 3.03. (4) At $\log P > 3.0$ only O-Miras (including
OH/IR Miras) have been observed. A similar situation exists in Our Galaxy
where there are no C-Miras with $\log P >3.0$ if one excepts one dubious
case with $\log P = 3.02$. The period distribution  of Galactic O-Miras (OH/IR 
stars) extends to longer than $\log P = 3.3$. 

The above results indicate a rather sharp change in the distribution and
composition of stars along the PL at $\log P \sim 3.0$. The discussion
of the previous section also leads to the conclusion that at this
period the Miras originate from stars of about 3 or $4 M_{\odot}$ . 
This is the approximate mass range above which HBB is expected to start.
Clearly if HBB does begin to operate at the mass of $\log P = 3.0$ Miras
it will explain the sudden change from essentially all C-Miras at slightly
shorter periods to all O-Miras at longer periods (and higher masses).
HBB adds an extra energy source to the star. At least to a first approximation
we would expect this to lead to an expansion of the star and hence to
a longer pulsation period. If such is the case then the gap just mentioned
leads to a measure of the extra energy produced. It is about 15 to 20 percent
of the total stellar energy production. 

If the stars with $\log P >3.0$ are HBB stars (together perhaps with some
post-HBB stars), 
how do we explain the above-PL
stars, some with lithium, which are found at $\log P \geq 2.7$?
The most likely solution is that these stars are overtone pulsators which have 
already started HBB and will evolve into O-Miras with $\log P \geq 3.0$.
A pointer in this direction is that one of these
above-PL  and lithium rich stars (IRAS 04496-6958)\cite{Whitelock-2003} 
is a possible member of a young cluster
LMC cluster HS33 \cite{Van Loon-2005}. If 
it is a member, it has an age of about 200Myr
and an initial mass of about $4 M_{\odot}$ , much larger than 
Miras on the PL at this $\log P$ (2.86). The above-PL stars have
relatively low amplitudes for their periods \cite{Whitelock-2003}
and humps on the rising branches of their light curves both in the
optical and the infrared \cite{Glass-2003} \cite{Rejkuba-2003}.
Ita et al. \cite{Ita-2004} have a sequence (C') of 
Mira-like
stars in their LMC $K- \log P$ plot 
which lies above and nearly parallel to the Mira sequence. If this
were plotted in $M_{bol}$ rather than $K$ it seems likely that our
above-PL stars would lie on an extrapolation of this sequence to longer
periods. 

Garci\'{a}-Hern\'{a}ndez et al. 
\cite{Garcia-Hernandez-2007} have examined a group of Galactic O-Miras
for strong lithium. Many of those with $\log P \sim2.6$ to 2.8 do show lithium
which they attribute to HBB (note that the longest period star in their fig 17
is WX Sgr, a supergiant not a Mira). If these Miras are indeed HBB stars
and belong with the above-PL stars then it hints that there are relatively few 
Galactic O-Miras
on the PL in this period range. A situation which, we have seen, seems to 
apply in the LMC. Some caution is however necessary. Their
shortest period lithium-rich star is  T Lep with $\log P = 2.57$.
A preliminary VERA parallax of the star reported at this meeting 
\cite{Nakagawa-2008}
places it on the PL. Thus it will be of too low a mass for HBB and
some other explanation of its lithium content is require. A likely
candidate is the cool bottom process \cite{Sackmann-1999}.

\bigskip
Hepful discussions with, and suggestions from, Patricia Whitelock
are gratefully acknowledged.


\end{document}